\documentclass[amsmath,twocolumn]{aastex63}

\usepackage{amsmath}
\usepackage{comment}

\usepackage{dcolumn}% Align table columns on decimal point
\usepackage{bm}% bold math

\usepackage{xcolor}

\newcommand{\etal}{\textit{et al}.}
\newcommand{\ie}{\textit{i}.\textit{e}.}
\newcommand{\eg}{\textit{e}.\textit{g}.}
\newcommand{\Ne}{$^{22}$Ne}
\newcommand{\Gammacrit}{$\Gamma_\mathrm{crit}$}

\submitjournal{ApJL}

\shorttitle{Neon Clustering and Phase Separation in WDs}
\shortauthors{Caplan, Horowitz, \& Cumming}
\graphicspath{{./}{figures/}}
\begin{document}

\title{Neon Cluster Formation and Phase Separation During White Dwarf Cooling} 

\author{M. E. Caplan}
 \email{mecapl1@ilstu.edu}
\affiliation{
 Illinois State University, Department of Physics, Normal, IL 61790 %\\
}
\author{C. J. Horowitz}\email{horowit@indiana.edu}
\affiliation{Center for Exploration of Energy and Matter and
                  Department of Physics, Indiana University,
                  Bloomington, IN 47405, USA}
\author[0000-0002-6335-0169]{A. Cumming}\email{andrew.cumming@mcgill.ca}
\affiliation{Department of Physics and McGill Space Institute, McGill University,
Montreal, QC H3A 2T8, Canada}




\begin{abstract}

Recent observations of Galactic white dwarfs (WDs) with Gaia suggest there is a population of massive crystallizing WDs exhibiting anomalous cooling -- the Q branch. While single-particle \Ne\ sedimentation has long been considered a possible heat source, recent work suggests that \Ne\ must separate into clusters, enhancing diffusion, in order for sedimentation to provide heating on the observed timescale. We show definitively that \Ne\ cannot separate to form clusters in C/O WDs using molecular dynamics simulations, and we further present a general C/O/Ne phase diagram showing that strong \Ne\ enrichment is not achievable for \Ne\ abundance $\lesssim 30\%$. We conclude that the anomalous heating cannot be due to \Ne\ cluster sedimentation and that Q branch WDs may have an unusual composition, possibly rich with heavier elements.

\end{abstract}



\section{Introduction} \label{sec:intro}

The Gaia space observatory has determined parallax distances to large numbers of Galactic stars \citep{GaiaNoAuthors}, which allow for unprecedented tests of white dwarf (WD) models and evolution. %Core crystallization, long predicted, is now resolved \cite{VanHorn1968,Tremblay2019}.  
Massive WDs have high central densities and strong gravities.  Furthermore, some of them may have formed via mergers \citep{WD_merger} and they may have interesting compositions.  Recently, \cite{Cheng_2019} found that the population of massive WD known as the `Q branch' appear to have an additional heat source that maintains a luminosity of order $10^{-3}\,L_\odot$ for Gyrs.  Latent heat from crystallization \citep{crystallization,Winget_2009,PhysRevLett.104.231101} and gravitational energy released from conventional $^{22}$Ne sedimentation \citep{Bildsten_2001,DeloyeBildsten2002,GarciaBerro2008,PhysRevE.82.066401} do not appear to be large enough to explain this luminosity \citep{Camisassa2020,Cheng_2019}.  
Heating from conventional electron capture and pycnonuclear (or density driven) fusion \citep{1969ApJ...155..183S,PhysRevC.74.035803,pycnoHorowitz} reactions appear to need even higher densities and may depend too strongly on the density and or temperature \citep{horowitz2020nuclear}.  There are many works discussing dark matter interactions in WD, see for example   \citep{PhysRevD.77.043515,PhysRevD.98.115027,PhysRevD.91.103514,PhysRevD.92.063007,PhysRevLett.115.141301,PhysRevD.100.043020}.  However, WD may be too small to capture enough dark matter for its annihilation to produce the necessary heat \citep{horowitz2020nuclear}.
%WD core crystallization has now been resolved observationally and motivates updated modeling of heat sources. 
%I) Introduction
%A) Massive WDs are interesting because of strong gravity and high central densities
%B) Cheng et al observe some massive WD that appear to need an additional heat source that maintains ~10^-3 L_sun for Gyrs.
%C) Electron capture, pycnonuclear fusion, and heating from dark matter all don’t work.

The gravitational potential of a WD is large.  Therefore, there is possibly enough energy available from sedimentation of neutron rich $^{22}$Ne to provide the necessary heating \citep{Bildsten2001}. However the expected diffusion constant for $^{22}$Ne in a C/O mixture \citep{hughto2010diffusion} is too small to allow enough sedimentation before the C/O mixture freezes. In addition, $^{22}$Ne sedimentation is significantly slowed down by C/O crystallization. Therefore, even  though sedimentation is a large enough energy source, in practice, sedimentation is likely slow and most of this energy may remain untapped by the time the star freezes \citep{DeloyeBildsten2002,GarciaBerro2008,PhysRevE.86.066413,PhysRevE.84.016401}.

It is possible that the $^{22}$Ne sedimentation rate is enhanced because of the formation of mesoscopic $^{22}$Ne clusters. Recently, \cite{2020arXiv200713669B} speculate that Ne phase separation produces mesoscopic clusters that could enhance the heating from conventional Ne sedimentation. Numerical calculations by \cite{BauerUnpub} consider this possibility in more detail, showing that clustering of $N$ nuclei enhances the downward drift rate proportional to $N^{2/3}$. This suggest that the appropriate diffusion timescale is achievable with long lived clusters of only few hundred to a few thousand \Ne\ ions, and is validated with MESA models.
%D) Sedimentation of neutron rich 22Ne could provide the necessary heat.  

In this letter we directly address this possibility with molecular dynamics (MD) simulations of Ne microcrystals in C/O mixtures and new calculations of the C/O/Ne phase diagram.  %We describe our MD formalism, present results for a number of simulations at different temperatures, discuss the implications and conclude.  
We find that our Ne microcrystals are unstable in C/O liquid mixtures.  Therefore enhanced heating from cluster formation or phase separation is unlikely to be important in conventional C/O/Ne mixtures. %and so some other heat source must be considered.   

%F) We directly test both possibilities with MD simulations.

%\lipsum[1-3]

%B) Why 49\% C, 49\% O, 2\% 22Ne mixture.

%truncated cuboctahedron

%Given the size of \Ne clusters proposed by Bauer and the almost exact nature of the interaction between ions in WDs, it is straightforward to evaluate whether these clusters are stable with molecular dynamics simulations. 

%\section{\label{sec:sim} Molecular Dynamics}

%\subsection{Simulations}

%A) Screened coulomb interactions and IUMD code

\section{\label{sec:sims}Molecular Dynamics Simulations}
%\textit{Simulations:}
IUMD is a CUDA-Fortran classical molecular dynamics code and has been extensively used to model astromaterials in WDs and neutron stars \citep{CaplanRMP}. In our model nuclei are fully ionized and treated as point particles of charge $Z_i$ which interact via a screened two-body potential %$V(r_{ij})=\frac{Z_i Z_j e^2}{r_{ij}} \exp(-r_{ij}/\lambda)$

\begin{equation}
V(r_{ij})=\frac{Z_i Z_j e^2}{r_{ij}} \exp(-r_{ij}/\lambda).
\label{eq.V}
\end{equation} 

\noindent with periodic separation $r_{ij}$ and screening length $\lambda^{-1}=2\alpha^{1/2}k_F/\pi^{1/2}$ using $k_F=(3\pi^2n_e)^{1/3}$. The screening assumes relativistic electrons, which is not the case in WDs. At $10^6$ g/cm$^3$, $k_F \approx 0.4$ MeV so we slightly overestimate $\lambda$; the true $\lambda$ is 21\% smaller. At $10^7$ g/cm$^3$, the true $\lambda$ is 7\% smaller. This is expected to have little impact on our MD (see \citealt{Hughto2011,PhysRevE.56.4671}). We use $\lambda/a = 2.816$ for our mixture (consistent with \citealt{hughto2010diffusion}) which forms a bcc lattice at low temperature \citep{PhysRevE.66.016404}. 
%The screening correction $kappa \equiv a/\lambda = 0.57$ is small

%These massive WDs are generally thought to have compositions by number abundance of 
The mixture we consider has number abundances $\vec{x} = (x_\mathrm{C}, x_\mathrm{O}, x_\mathrm{Ne}) \approx (0.49, 0.49, 0.02)$ and is well motivated astrophysically.
The $^{4}\mathrm{He}(2\alpha,\gamma)^{12}\mathrm{C}$ and $^{12}\mathrm{C}(\alpha,\gamma)^{16}\mathrm{O}$ reactions set $x_\mathrm{C}/x_\mathrm{O}\approx1$, with order 10\% variation due to reaction rates and WD mass \citep{lauffer2018new}. During helium burning the remaining CNO elements are thought to burn via $^{14}\mathrm{N}(\alpha,\gamma)^{18}\mathrm{F}(\beta^+)^{18}\mathrm{O}(\alpha,\gamma)^{22}\mathrm{Ne}$; assuming solar metallicity, $x_\mathrm{Ne}\approx 0.02$ in the WD \citep{bildsten2001gravitational}. %Constraining the composition by the presence of absence of heat sources in WDs can therefore have important implications for these reaction rates. 

\begin{figure}[t!]
\centering  %left bottom top right)
\includegraphics[trim=180 270  180 270,clip,width=0.47\textwidth]{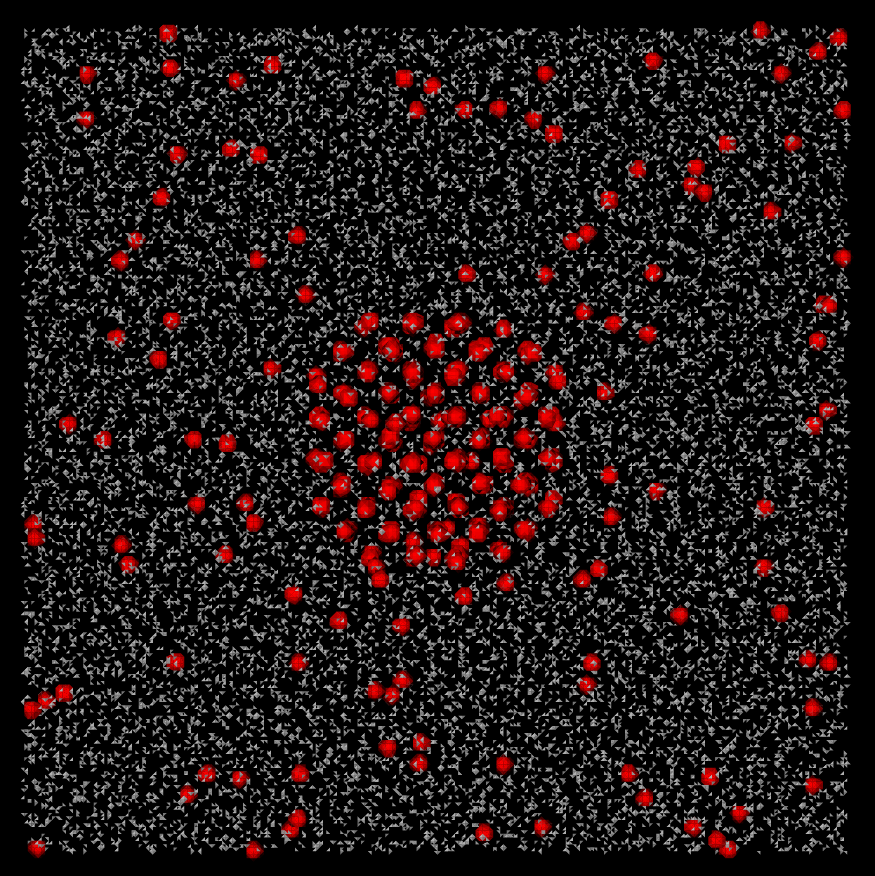}
\caption{\label{fig:MD_init} (Color online) Initial conditions, \Ne\ (red) is concentrated in a bcc microcrystal  in the center of the volume and is surrounded by a C/O (white) liquid with trace \Ne. The configuration is cubic, an orthographic projection is used for clarity.}	
\end{figure}

The crystallization of a one-component plasma (OCP) depends on the dimensionless parameter $\Gamma_{i} = Z_i^{5/3}e^2 / a_e k_B T$ with ion charge $Z_i$, electron separation $a_e = (3/4 \pi n_e)$, and thermal energy $k_B T$. The OCP is solid (liquid) above (below) $\Gamma_{\mathrm{crit}} \approx 175$ \citep{PhysRevE.62.8554}. Screening may raise \Gammacrit\ to about 178 \citep{PhysRevE.66.016404}.
Averaging over mixture components for the multi-component plasma (MCP) gives $\Gamma_{\mathrm{MCP}} = \langle Z \rangle^{5/3}e^2 / a_e k_B T$; crystallization is more complicated and sensitive to the exact mixture. A binary C/O mixture has  $\Gamma_{\mathrm{MCP}} \approx \Gamma_\mathrm{crit} \approx230$, while C/O/Ne mixtures with $x_\mathrm{Ne} = 0.2$ have been found with MD to have $275 \lesssim \Gamma_\mathrm{crit} \lesssim 300$ \citep{Hughto2011}. If Ne clusters are unstable then we expect that the \Gammacrit\ we find will be closer to 300 than 230, as the region of the phase diagram probed will be that of high Ne concentration at the C/O liquid-Ne cluster surface. %Furthermore, the crystal is small and we use short run times. %, we use small run times, and the initial condition is likely out of equilibrium. 

%\Gammacrit\ for $\vec{x}=(0.49, 0.49, 0.02)$. 

%The screening correction depends on the dimensionless ratio of the characteristic ion spacing and the screening length, $kappa \equiv n^{-1/3}/\lambda = 0.57$. The screening correction to \Gammacrit\ is uncertain at small $\kappa$; the corrections compiled by \citealt{PhysRevE.66.016404} suggest correction factors of $e^\kappa / (1 + \kappa + 1/2 \kappa^2) \approx 1.02$ or $1+0.1\kappa = 1.05$. 

%While exact abundances are dependent on 

%We expect the ratio of $x_C/x_O \approx 1$

%C)Preparing initial 22Ne microcrystal

Our MD mixture contains 8000 $^{12}$C, 8000 $^{12}$O, and 384 \Ne\ ions ($N=16384=2^{14}$, for GPU threading). %If \Ne\ clusters form then it occurs via spontaneous nucleation of a small crystallite which then grows by adsorption. Such a cluster may then be approximately spherical; sharp edges as in cubic nanoparticles may be strongly disfavored due to the high surface energy. 
%D) Initial conditions
Our initial configuration is shown in Fig. \ref{fig:MD_init} and consists of a neon microcrystal embedded in a C/O liquid in a cubic volume with periodic boundaries.  The neon crystal is prepared by trimming ions from the edges and corners of a cubic bcc crystal to produce a truncated cuboctahedron containing 253 ions, $6a$ on its longest diameter. 
%This is a standard technique for studying terrestrial nanoparticles in the literature. 
The C/O fluid and the remaining \Ne\ have random initial positions around the microcrystal; if the microcrystal is stable we expect it to grow by adsorption of the remaining 131 \Ne.

%The neon crystallite we consider is slightly smaller than the size proposed by Bauer \etal\, $N\approx 300-3000$. Our smaller simulations have the advantage of running quickly and allowing us to run for a large number of timesteps. Furthermore, the arguments of Bauer suggest that these mesoscopic clusters form spontaneously so if they are stable we expect that our cluster should grow by adsorption of the remaining \Ne\ from the liquid; if the \Ne\ crystallite is unstable on MD timescales then this implies that it is not possible for MD clusters to grow to the size proposed by Bauer. 
%E) MD runs at constant temperature

We run isothermal simulations to resolve \Gammacrit\ and study the stability of the neon cluster at a range of temperatures. Constant temperature is approximately achieved by rescaling the velocities to a Maxwell-Boltzmann distribution with the desired temperature every 100 timesteps. Our simulations therefore do not conserve energy; instead, in the long time limit our simulations trend toward equilibrium so $E(t)$ allows us to resolve melting or freezing (\eg\ heats of fusion). Our simulations differ from past work (\eg\ \citealt{Hughto2011}) and our asymptotic states may not be true equilibrium (\ie\ they may be superheated/cooled) because we only run as long as needed to verify stability or instability of the neon cluster.

\begin{figure}[t!]  %max width = 0.2375
\centering  %left bottom top right)
\includegraphics[trim=170 255  170 255,clip,width=0.23\textwidth]{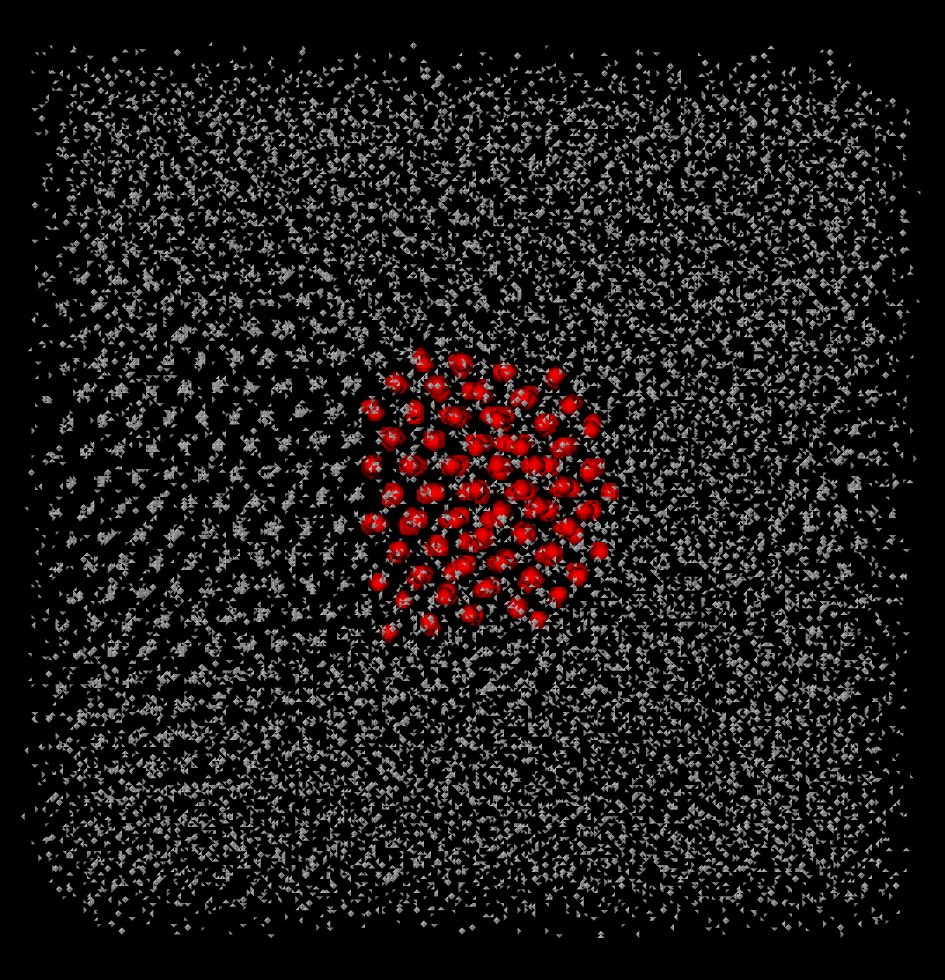}
\includegraphics[trim=170 255  170 255,clip,width=0.23\textwidth]{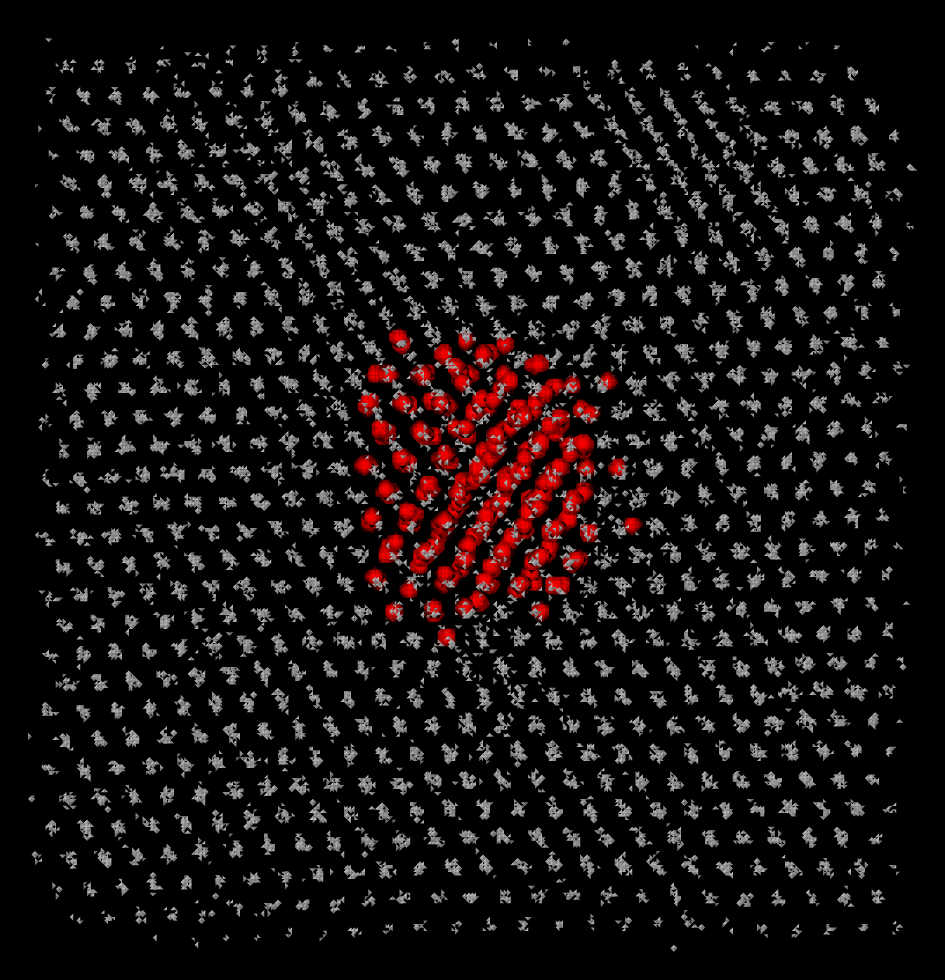}
\includegraphics[trim=180 270  180 270,clip,width=0.23\textwidth]{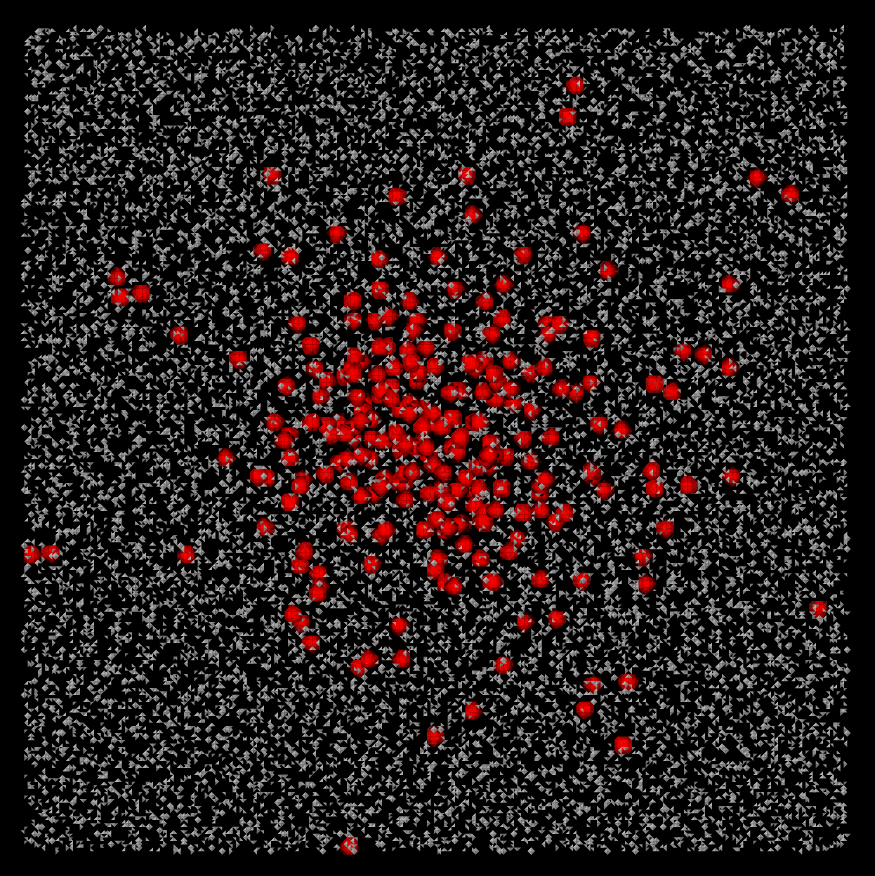}
\includegraphics[trim=180 270  180 270,clip,width=0.23\textwidth]{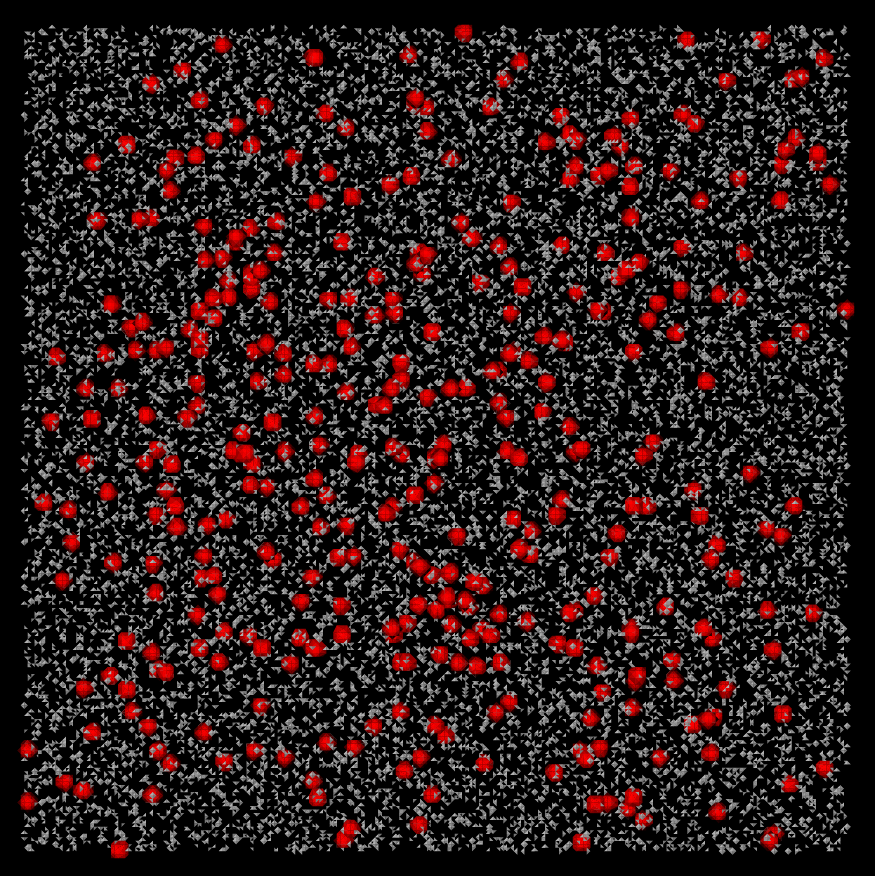}
\caption{\label{fig:MD_equil} (Color online) Intermediate (left) and final (right)  configurations of simulations at $\Gamma = 289$ (top, solid) and $\Gamma=287$ (bottom, liquid). Colors as in Fig. \ref{fig:MD_init}; for clarity we only show the cluster \Ne\ in red.}
\end{figure}

\begin{table}[]
\caption{\label{tab:md} Summary of MD runs.}
\begin{tabular*}{0.47\textwidth}{c c c c c} \hline \hline
$\Gamma$ & $(\Gamma_\mathrm{C},\Gamma_\mathrm{O},\Gamma_\mathrm{Ne})$ & & Outcome                   &  \\ \hline
296    & (222,359,520)                & & C/O Crystallization around Ne &    \\ 
291    & (219,353,513)                & & C/O Crystallization around Ne &    \\
289    & (217,351,509)                & & C/O Crystallization around Ne &    \\
287    & (215,348,505)                & & Ne cluster dissolves into C/O &    \\
285    & (214,346,501)                & & Ne cluster dissolves into C/O &    \\
244    & (183,296,429)                & & Ne cluster dissolves into C/O &    \\ \hline \hline
\end{tabular*}
\end{table}

In Tab. \ref{tab:md} we list isothermal simulations run using the initial configuration shown in Fig. \ref{fig:MD_init}. These simulations were run for between $10^5$ and $10^6$ MD timesteps with $dt = 1/18 \omega_p$ with ion plasma frequency $\omega_p = ( 4 \pi e^2 \langle Z \rangle^2 n / \langle M \rangle )^{1/2}$. %( \Sigma_i Z_i^2 4 \pi e^2 x_i n / m_i)^{1/2}$). 
We clearly resolve a first order transition between $287 < \Gamma < 289$. At $\Gamma \geq 289$ the C/O liquid is supercooled and quickly begins crystallizing, first nucleating around the Ne cluster (Fig. \ref{fig:MD_equil} top left) before growing to fill the volume (top right).  At $\Gamma \leq 287$ we find that the Ne microcrystal dissolves into the liquid (bottom left) and in the long time limit is fully mixed in the volume (bottom right). At $\Gamma=244$ the microcrystal immediately melts to a liquid and mixes, while at higher $\Gamma$ the cluster seems to sublimate as the core remains solid while ions slowly desorb into the liquid. % retains some solid while the outer layers sublimate. 
When the configuration with solid C/O is evolved for longer times we also resolve slow diffusion on the lattice which seems to disperse the Ne enriched core. We have thus shown that our initial condition is unstable, suggesting that \Ne\ microcrystals are not expected in C/O WDs.  

Our $\Gamma_\mathrm{crit} \approx 288$ is high compared to past work for C/O/Ne mixtures, which find $\Gamma_\mathrm{crit} \approx 230$ (see Tab. 1 in \citealt{PhysRevE.86.066413}). Our system may be finding the melting point of a very Ne rich system. Furthermore, the simulation does not have enough time for diffusion to bring the composition of the solid phase into equilibrium with the composition of the liquid phase. Both finite size and finite time effects may be important in comparing to an equilibrium phase diagram (computed below).

\begin{figure}[t!]
\centering
\includegraphics[trim=55 145 55 145,clip,width=0.3375\textwidth]{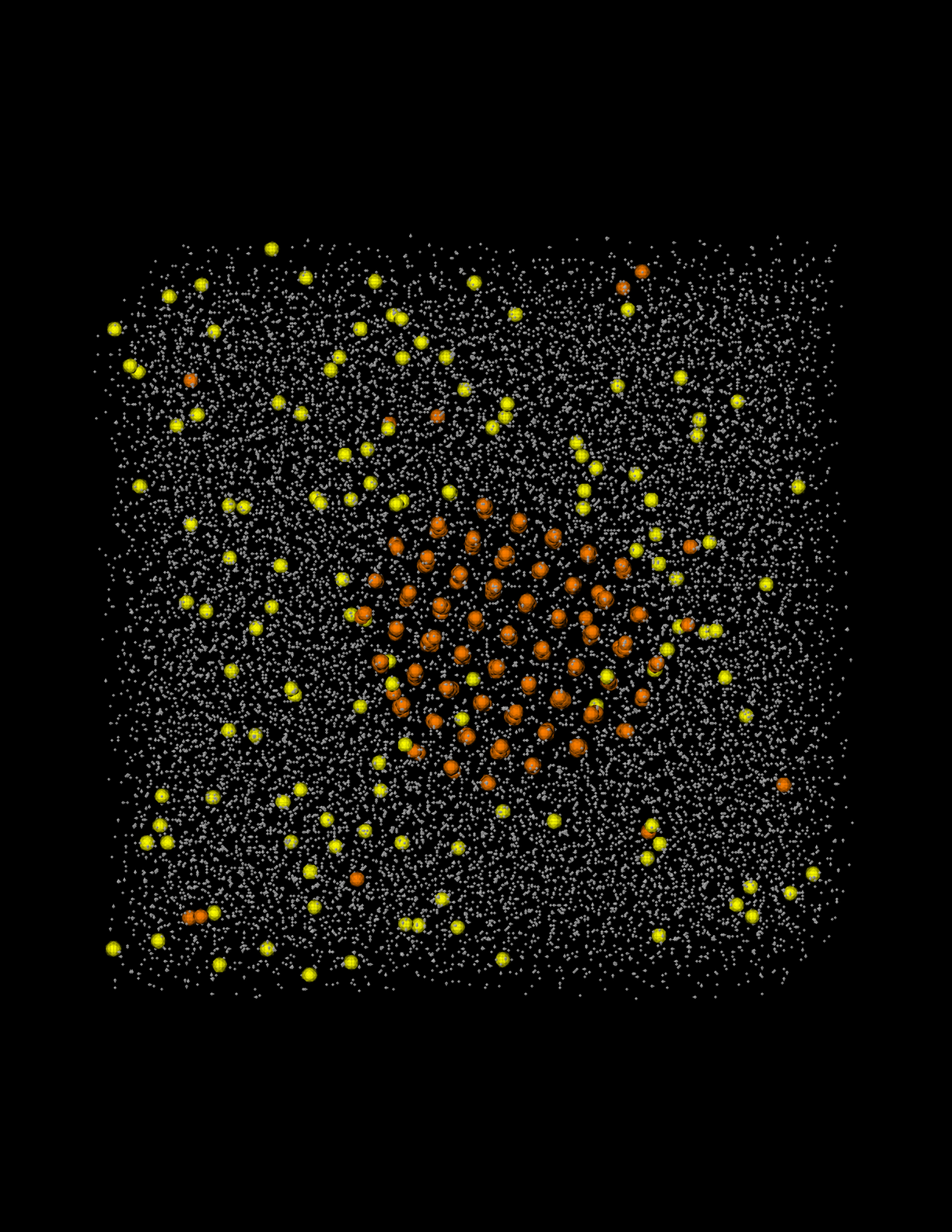}
\caption{\label{fig:Fe} (Color online) Evolved Fe microcrystal at $\Gamma = 289$ ($\Gamma_\mathrm{C}=187$, $\Gamma_\mathrm{O}=301$, $\Gamma_\mathrm{Fe}=2149$). Some small amount of Fe initially in the microcrystal (orange) escapes into the background or rearranges at edges, while a similar amount of background Fe (yellow) is captured on the surface.} 
\end{figure}

We now consider what impurities might form stable microcrystals. Generally speaking, stronger separation is observed in mixtures with greater contrast in charge $Z$. While $^{23}$Na or an isotope of Mg may be present in comparable abundances to \Ne\, they may not have large enough $Z$ to strongly separate. Simulations with Mg in place of the Ne at $\Gamma=290$ also show a slow sublimation of the cluster into the liquid, though on slightly longer timescales than the Ne. 
%/N/slate/mecaplan/CLUSTER_22NE/MAGNESIUM/RUN4
In simulations with Fe run at $\Gamma =$ 240, 262, and 289 the microcrystal persists after a few times $10^7$ timesteps. A configuration evolved for $4.1\times10^7$ timesteps at $\Gamma = 289$ is shown in Fig. \ref{fig:Fe}. 
In all three simulations the microcrystal shows exchange with the Fe in the background and evolution in morphology. Ions on raised facets seem more likely to escape or migrate to adjacent faces to produce larger smooth surfaces, possibly an octahedron (stability, growth, and diffusion of various microcrystal morphologies may be of interest to future authors). Thus, while \Ne\ does not form stable microcrystals, higher $Z$ impurities are viable candidates for phase separation and clustering.

%... at $\Gamma < \Gamma_\mathrm{crit}$ the neon cluster is a superheated s ...

%Meh, probably cut this next paragraph
%The CO phase diagram and MD of C/O/Ne suggests that the mixture is unstable in a region between about 0.95\Gammacrit\ to 1.05\Gammacrit\, so the exact  \Gammacrit\ we resolve by MD may be influenced by finite size and time effects as well as the energy flux to the system from velocity rescaling. We verify phase coexistence at a range of temperatures by simulating a partially-frozen configuration generated from a the simulations described above.

%^, but we expect it to be in a region of approximately $240 < \Gamma < 300$. 

%\subsection{\label{sec:res}Results}

%A) The whole system freezes for large Gamma or

%B) Ne microcrystal evaporates for smaller Gamma

%C) No conditions where Ne microcrystal is stable in a C/O liquid.

%            D) Pretty pictures!

%E) Phase diagram

\begin{figure}[t!]
\centering
\includegraphics[trim=0 0 0 0,clip,width=0.4\textwidth]{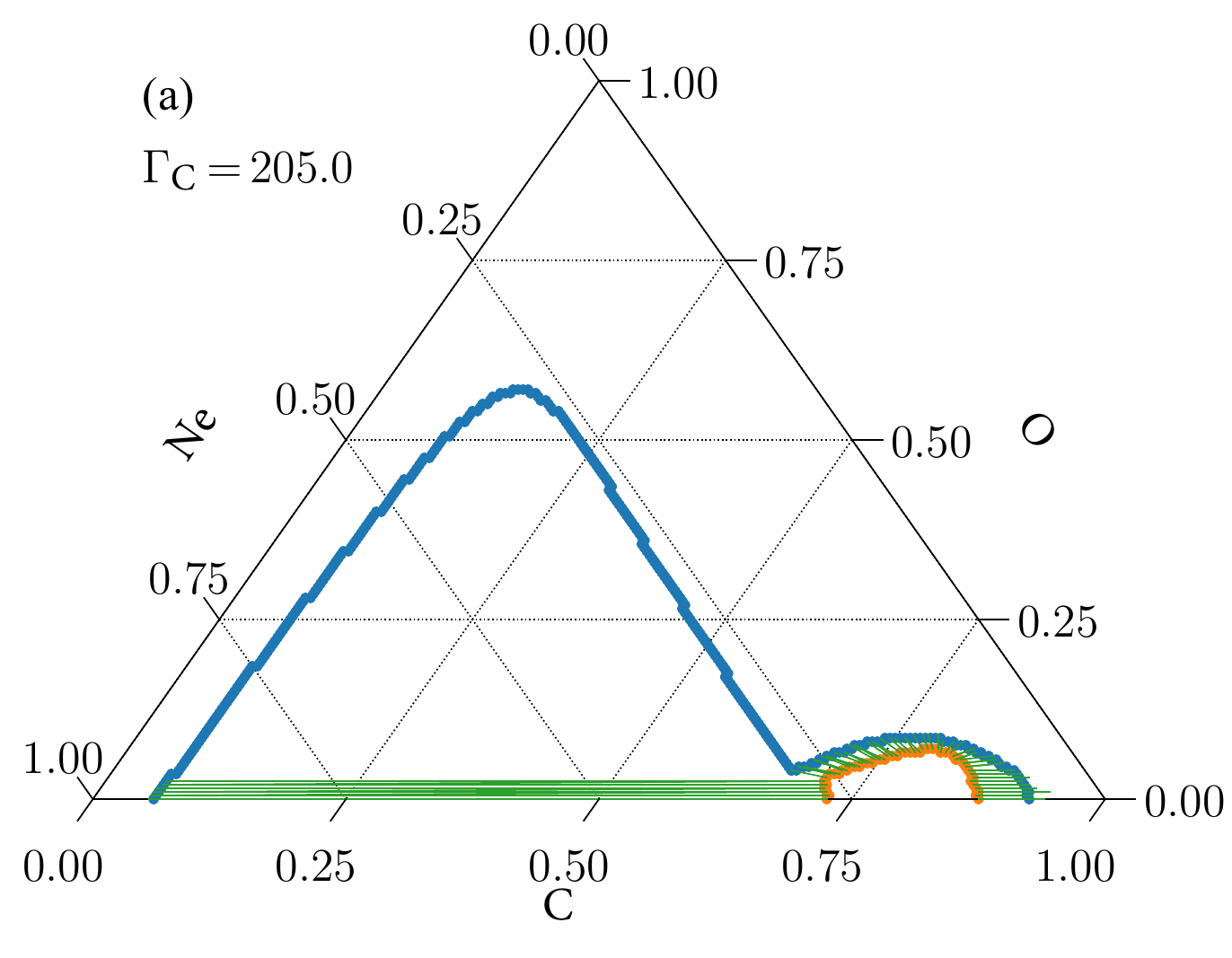}
\includegraphics[trim=0 0 0 0,clip,width=0.4\textwidth]{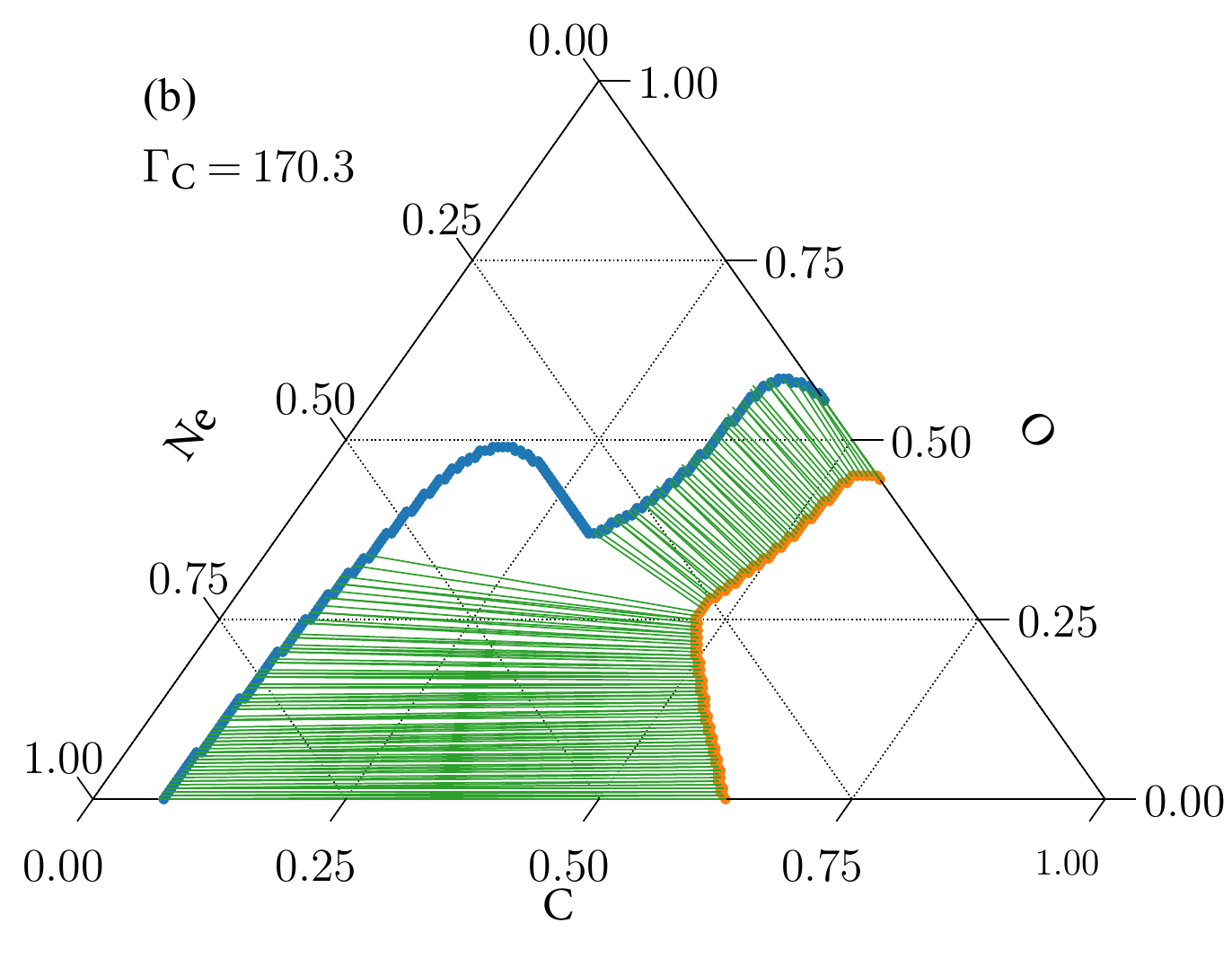}
\includegraphics[trim=0 0 0 0,clip,width=0.4\textwidth]{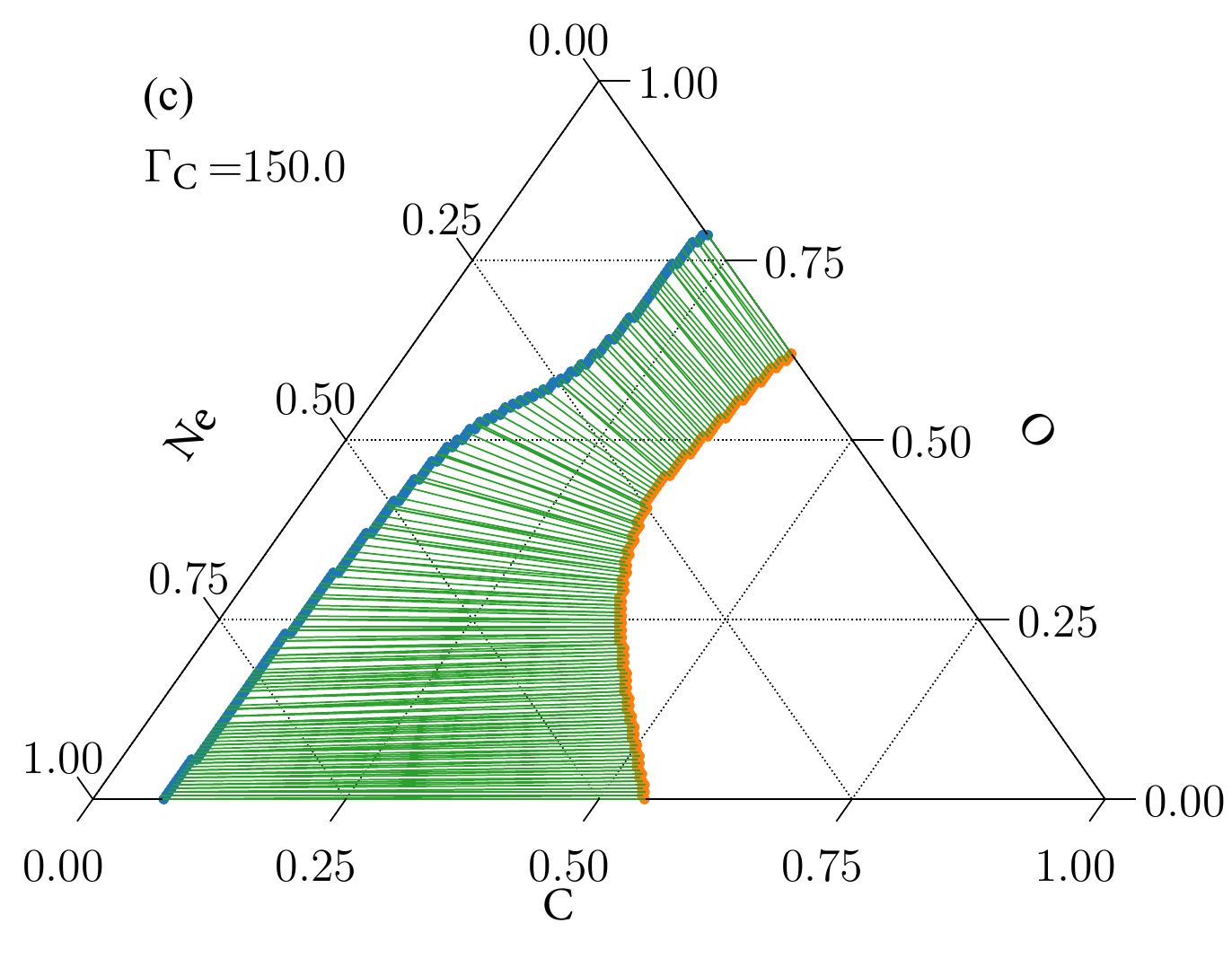}
\caption{\label{fig:PD} 
(Color online) C/O/Ne phase diagram. The liquidus (orange) and solidus (blue) are connected by tie-lines (green) showing solid-liquid equilibrium. 
To find a given composition $\vec{x}=(x_\mathrm{C}, x_\mathrm{O}, x_\mathrm{Ne})$ in the phase diagram lines of constant $x_i$ are projected from the slope of the tick mark on the relevant axis. For example, pure Ne is found in the bottom left corner while two-component CO mixtures are found along the right axis, so our $\vec{x} = (0.49, 0.49, 0.02)$ mixture lies at a point near the middle of the right axis.}
\end{figure}

\section{\label{sec:pd}C-O-Ne Phase Diagram}
%\textit{C/O/Ne Phase Diagram:}
Although mixtures with $x_\mathrm{Ne}\approx 0.02$ cannot form stable microcrystals, mixtures with larger Ne abundances could. Therefore, we compute the ternary phase diagram to determine what Ne abundance may be required for such strong phase separation. We use the code\footnote{The code is available at \url{https://github.com/andrewcumming/phase_diagram_3CP}.
}
developed in \cite{Caplan2018} which implements the semi-analytic method of \cite{Medin2010}. This method identifies pairs of points on the minimum free energy surfaces which share a tangent plane (\ie\ the double tangent construction); these points correspond to coexisting solid and liquid compositions. Compositions lying on the tangent line connecting them are therefore unstable and phase separate. 
\begin{comment}
We use liquid free energy
\begin{equation}
\begin{split}
f_l^{ \rm{MCP} } & (\Gamma_1,x_1,\Gamma_2,x_2,\Gamma_3,x_3)  \simeq  \\ 
& \sum_{i=1}^3 x_i \left[ f_l^{ \rm{OCP} }(\Gamma_i) + \ln\left(x_i \frac{Z_i}{\langle Z \rangle}\right)\right]
\end{split}
\end{equation}
\noindent and solid free energy
\begin{equation}
\begin{split}
f_s^{ \rm{MCP} } & (\Gamma_1,x_1,\Gamma_2,x_2,\Gamma_3,x_3)  \simeq  \\ 
& \sum_{i=1}^3 x_i \left[ f_s^{ \rm{OCP} }(\Gamma_i) + \ln\left(x_i \frac{Z_i}{\langle Z \rangle}\right)\right] \\
& + \delta f_s (\Gamma_1,x_1,\Gamma_2,x_2,\Gamma_3,x_3)
\end{split}
\end{equation}
using ion abundance $x_i$ with charge $Z_i$ (and $\Gamma_i$), the deviation from linear mixing for solids $\delta f_s$, and the solid and liquid OCP free energies $f_s ^{ \rm{OCP}}$ and $f_l^{ \rm{OCP}}$. 
\end{comment}
For a detailed discussion see \cite{Medin2010,Caplan2018}.

In Fig. \ref{fig:PD} we show C/O/Ne phase diagrams for three $\Gamma$. We report temperature in $\Gamma_\mathrm{C} \propto 1/T$. Orange liquidus points are connected to corresponding blue solidus by green tie lines corresponding to the tangent in free energy. Below (above) the orange (blue) curve is stable liquid (solid), while the green tie lines span the unstable region. 

At the lowest temperature (Fig. \ref{fig:PD}a), it is possible to form solid particles that are substantially enriched in Ne. The phase diagram shows solid-solid coexistence with strongly Ne enriched mixtures, where both the Ne-enriched and Ne-depleted crystals show $x_\mathrm{Ne} > 0.3$ (for readability we exclude solid-solid tie lines). We see the emergence of stable liquid near the C/Ne axis (bottom), found in the white region under the liquidus near $x_\mathrm{C}\approx 0.75$. Along the C/Ne axis we see solid-liquid phase coexistence with strong separation; given $Z_\mathrm{Ne} / Z_\mathrm{C} = 1.66$ we resolve eutectic separation. 

At intermediate temperature (Fig. \ref{fig:PD}b) we reach \Gammacrit\ for our mixture, $\vec{x}= (0.49, 0.49, 0.02)$, which is found near the middle of the RHS axis. In the unstable region we find separation largely consistent with the known C/O phase diagram from MD (see Fig. 5 in \citealt{PhysRevE.86.066413}). Separation is approximately independent of Ne fraction for $x_\mathrm{Ne} \lesssim 0.3$. Observe that the coexistence lines are approximately parallel to the C/O axis, which implies they fall on lines of constant Ne. Therefore, the Ne fraction in the solid and liquid are nearly equal which suggests that any C/O/Ne mixture with  $x_\mathrm{Ne} \lesssim 0.3$ is incapable of separating to form any Ne rich components which could then preferentially settle out. This is qualitatively consistent with past MD (see \citealt{Hughto2011}) which studied mixtures of $0.02 \leq x_\mathrm{Ne} \leq 0.20$.

We also resolve a small set of mixtures which undergo three-component eutectic separation at intermediate $\Gamma$, similar to the mixtures with large $Z_2/Z_1$ and $Z_3/Z_1$ studied by \cite{Caplan2018}. The central `wedge' between the break in tie lines contains unstable mixtures which do not fall on a single coexistence line so they cannot separate into a single solid and single liquid composition, but they can separate by forming appropriate amounts of the two solids and one liquid at the corners of this region.

At high temperature (Fig. \ref{fig:PD}c) the solidus and liquidus curves are continuous and the region for the eutectic separation has closed. Though mixtures with $x_\mathrm{Ne} \approx 0.30$ now show weak enhancement in solidus Ne, those with $x_\mathrm{Ne} \lesssim 0.30$ are still largely consistent with two-component C/O separation without any Ne enrichment or depletion. 

Even if there is factor of 2 variation in $x_\mathrm{O}/x_\mathrm{C}$ it only translates our mixture parallel to the O axis, which does not move our mixture into a region where it achieves strong Ne purification at any $\Gamma$. We conclude that the behavior of \Ne\ microcrystals in MD is fully consistent with the known C/O/Ne phase diagram, and that no mesoscopic effects exist that make \Ne\ cluster formation likely. 

As past MD has only studied phase coexistence up to 20\% Ne, we also perform a few MD simulations to validate the behavior around 30\% Ne. Similar to \citealt{PhysRevE.86.066413,Caplan2018}, configurations are prepared by joining a cubic bcc crystal on one face with a cubic volume of liquid (as in Fig. 1 in \citealt{PhysRevE.86.066413} and Fig. 1 in \citealt{Caplan2018}). Compositions for the solid and liquid are chosen to approximately match predictions from the phase diagram at $\Gamma_\mathrm{C}=183$, allowing us to verify the strong break in the center of Fig. \ref{fig:PD}b. Our first mixture has low Ne, $\vec{x} = (0.4,0,3,0.2)$, which separates to a solid $\vec{x}_s \approx (0.4,0.3,0.3)$ and a liquid $\vec{x}_l \approx (0.6,0.2,0.2)$. Our second mixture has high Ne, $\vec{x} =  (0.32,0,15,0.53)$, and strongly separates to $\vec{x}_s \approx  (0.1, 0.15, 0.75)$ and $\vec{x}_l \approx  (0.55, 0.15, 0.3)$. These simulations were evolved for $10^7$ timesteps with little observed evolution. Runs varying $\Gamma_\mathrm{C}$ up and down respectively find quenching of diffusion in the liquid and melting of the crystal, suggesting these temperature variations have moved our mixtures into the region of stable liquid and stable solid. Taken together, these simulations have qualitative agreement with our phase diagram, and we conclude that the separation behavior near 30\% Ne concentration is likely physical. Future work may be interested in performing a more thorough survey at high Ne concentration with MD, though this may have limited astrophysical relevance.

\section{\label{sec:dis}Discussion}
%\textit{Discussion:} 
We find that $^{22}$Ne microcrystals are always unstable in a C/O liquid.  Either the temperature is high enough that the crystal melts and the Ne dissolves into the liquid, or the whole system including the C/O mixture freezes.  Note that even at temperatures below the melting point of pure Ne, but above the C/O melting point, a large entropy of mixing causes the small concentration of Ne to dissolve into the bulk liquid. The C/O/Ne phase diagram suggests that very much more Ne is necessary before it phase separates.  One needs not 2\% but $\approx$ 30\% or more.  As a result, a conventional C/O WD with $n_\mathrm{Ne} =0.02$ is not expected to form stable neon clusters with enhanced sedimentation.  %Even though sedimentation is a large enough energy source, in practice, sedimentation is likely slow and most of this energy may remain untapped by the time the star freezes. 
In summary, we find that there are no conditions where a \Ne-enriched cluster is stable in a C/O WD, and therefore, enhanced diffusion of \Ne\ cannot explain the Q branch.

What compositions could then explain the heating that Cheng \etal~infer?  As seen in our phase diagram, unless the C/O ratio or \Ne\ abundance is tuned to extremes we don't expect strong \Ne\ separation, so we suggest that Q branch WDs may have an anomalous composition. For example, \cite{Camisassa2020} have suggested that $x_\mathrm{Ne}=0.06$ can provide heating on the desired timescale considering only single-particle diffusive settling rather than clusters. 
Another possibility is $\approx1\%$ abundance of another impurity, besides $^{22}$Ne, with an even larger charge $Z$ which would allow it to phase separate even when Ne does not. %that has a charge $Z$ more different (then Ne) from the average charge of the bulk mixture.  This would allow it to phase separate, even when Ne does not.  
 %Possibly $Z=12$ Mg, or larger $Z$, will separate out of a C/O mixture. 
This impurity would need to be neutron rich ($Z/A<0.5$) to be a sedimentary heat source and have an abundance of a percent or more for there to be enough gravitational energy available. Our MD with a high purity microcrystal shows that $Z=11$ Na and $Z=12$ Mg should not strongly separate in a C/O mixture. Phase diagrams of C/O/Na and C/O/Mg mixtures produced using our semi-analytic method (omitted for length) are similar to the C/O/Ne in that they do not separate to form a solid enriched in the high $Z$ impurity when it is only abundant at the percent level, so isotopes such as $^{23}$Na or $^{26}$Mg are poor candidates for clustering. 

%However, nuclides such as $^{23}$Na or $^{26}$Mg may still be viable candidates if phase separation can even weakly enhance their concentration in the solid phase in mixtures with low abundances, motivating work on these previously less studied phase diagrams. Future work could study the behaviour of $^{23}$Na or $^{26}$Mg in C/O mixtures and also investigate the four component C/O/Ne/Na or C/O/Ne/Mg phase diagram. % phase diagram for a star that has undergone some C burning.

Iron-group elements provide another possibility, as we readily observe long lived Fe microcrystals in MD. As $\Gamma_\mathrm{Fe} \approx 10 \Gamma_\mathrm{C/O}$, Fe in C/O will phase separate and likely does not require fine tuning of the mixture. While sedimentation of 0.1\% Fe by mass may produce some notable heating, if some astrophysical process enriches Q branch WDs up to $\approx 1\%$ mass fraction then settling out of Fe could provide heating for several Gyr as an otherwise conventional C/O WD cools. Thus, this work motivates including Fe in WD cooling models. This will require new phase diagrams of Fe and a survey with MD of the clustering and the characteristic sizes of Fe clusters, which will be the subject of future work.
%It may be interesting in future work to include Fe in WD cooling models, though new phase diagrams and MD may first be required to evaluate the plausibility of clustering and the characteristic sizes.  

% iron 
%If the $^{56}$Fe abundance is not $\approx 0.1\%$ but of order 1\% instead then the settling out of Fe could provide heating for several Gyr as an otherwise conventional C/O WD cools.  Note that sedimentation of even 0.1\% Fe may produce notable heating. 

%\lipsum[1-3]

\acknowledgments
     We thank S.~Cheng and E.~Bauer for helpful discussions. CH's research was supported in part by US Department of Energy Office of Science grants DE-FG02-87ER40365 and DE-SC0018083. The authors acknowledge the Indiana University Pervasive Technology Institute for providing supercomputing and database, storage resources that have contributed to the research results reported within this paper. This research was supported in part by Lilly Endowment, Inc., through its support for the Indiana University Pervasive Technology Institute. AC is supported by an NSERC Discovery Grant, and is a member of the Centre de recherche en astrophysique du Québec (CRAQ).

%\bibliography{apsbib}
\providecommand{\noopsort}[1]{}\providecommand{\singleletter}[1]{#1}%

\bibliographystyle{aasjournal}



\end{document}